\documentclass[a4paper,showpacs,showkeys,nofootinbib,aps]{revtex4}
\usepackage{epsfig}
\begin{document}
\boldmath
\title{
\vskip-3cm{\baselineskip14pt
 \centerline{\normalsize DESY~08-081 \hfill ISSN 0418-9833}
 \centerline{\normalsize June 2008\hfill}} \vskip1.5cm
On the Perturbative Stability of the QCD Predictions\\
for the Ratio $R=F_L/F_T$ in Heavy-Quark Leptoproduction}
\unboldmath
\author{N.Ya.~Ivanov}
 \email{nikiv@mail.yerphi.am}
\affiliation{Yerevan Physics Institute, Alikhanian Br.~2, 375036 Yerevan, Armenia}
\author{B.A.~Kniehl}
 \email{kniehl@desy.de}
\affiliation{{II.} Institut f\"{u}r Theoretische Physik, Universit\"{a}t Hamburg, Luruper
Chaussee 149, 22761 Hamburg, Germany}

\begin{abstract}
\noindent We analyze the perturbative and parametric stability of the QCD predictions for
the Callan-Gross ratio $R(x,Q^2)=F_L/F_T$ in heavy-quark leptoproduction.
We consider the radiative corrections to the dominant photon-gluon fusion mechanism. In
various kinematic regions, the following contributions are investigated: exact NLO
results at low and moderate $Q^2\lesssim m^2$, asymptotic NLO predictions at high $Q^2\gg
m^2$, and both NLO and NNLO soft-gluon (or threshold) corrections at large Bjorken $x$.
Our analysis shows that large radiative corrections to the structure functions
$F_T(x,Q^2)$ and $F_L(x,Q^2)$ cancel each other in their ratio $R(x,Q^2)$ with good
accuracy. As a result, the NLO contributions to the Callan-Gross ratio are less than $10\%$ 
in a wide region of the variables $x$ and $Q^2$. 
We provide compact LO predictions for $R(x,Q^2)$ in the case of low $x\ll 1$. A simple 
formula connecting the high-energy behavior of the Callan-Gross ratio and low-$x$ asymptotics 
of the gluon density is derived. It is shown that the obtained hadron-level predictions for 
$R(x\to 0,Q^2)$ are stable under the DGLAP evolution of the gluon distribution function. 
Our analytic results simplify the extraction of the structure functions $F_2^c(x,Q^2)$ 
and $F_2^b(x,Q^2)$ from measurements of the corresponding reduced cross sections, 
in particular at DESY HERA.
\end{abstract}
 \maketitle

\section{Introduction}

In the framework of perturbative quantum chromodynamics (QCD), the basic spin-averaged
characteristics of heavy-flavor hadro-
\cite{Nason-D-E-1}, photo- \cite{Ellis-Nason} and electro-production \cite{LRSN} are known
exactly up to the
next-to-leading order (NLO). Although these explicit results are widely used at present
for a phenomenological description of available data (for reviews, see
Refs.~\cite{FMNR-rev, R-Vogt}), the key question remains open: How to test the
applicability of QCD at fixed order to heavy-quark production? The basic theoretical
problem is that the NLO corrections are sizeable; they increase the leading-order (LO)
predictions for both charm and bottom production cross sections by approximately a factor
of two. Moreover, soft-gluon resummation of the threshold Sudakov logarithms indicates
that higher-order contributions can also be substantial. (For reviews, see
Refs.~\cite{Laenen-Moch,kid1}.) On the other hand, perturbative instability leads
to a high sensitivity of the theoretical calculations to standard uncertainties in the
input QCD parameters. For this reason, it is difficult to compare pQCD results for
spin-averaged cross sections with experimental data directly, without additional
assumptions. The total uncertainties associated with the unknown values of the heavy-quark
mass, $m$, the factorization and renormalization scales, $\mu _{F}$ and $\mu _{R}$, the
asymptotic scale parameter $\Lambda_{\mathrm{QCD}}$ and the parton distribution functions (PDFs)
are so large that one can
only estimate the order of magnitude of the pQCD predictions for charm production cross
sections in the entire energy range from the fixed-target experiments
\cite{Mangano-N-R} to the RHIC collider \cite{R-Vogt}.

Since these production cross sections have such slowly converging
perturbative expansions, it is of special interest
to study those observables that are well-defined in pQCD. A nontrivial example of such an
observable was proposed in Refs.~\cite{we1,we2,we3,we4,we5}, where the azimuthal
$\cos(2\varphi)$ asymmetry in heavy-quark photo- and leptoproduction was
analyzed.\footnote{Well-known examples include the shapes of differential cross sections
of heavy-flavor production, which are sufficiently stable under radiative
corrections.}$^{,}$\footnote{Note also the recent paper \cite{Almeida-S-V}, where the
perturbative stability of the QCD predictions for the charge asymmetry in top-quark
hadroproduction has been observed.} In particular, the Born-level results were
considered \cite{we1} and the NLO soft-gluon corrections to the basic mechanism,
photon-gluon fusion, were calculated \cite{we2}. It was shown that, contrary to
the production cross sections, the azimuthal asymmetry in heavy-flavor photo- and
leptoproduction is quantitatively well defined in pQCD: the contribution of the dominant
photon-gluon fusion mechanism to the asymmetry is stable, both parametrically and
perturbatively. Therefore, measurements of this asymmetry should provide a useful test of
pQCD. As was shown in Ref.~\cite{we3}, the azimuthal asymmetry in open charm
photoproduction could be measured with an accuracy of about ten percent in the
approved E160/E161 experiments at SLAC \cite{E161} using the inclusive spectra of
secondary (decay) leptons.

In Ref.~\cite{we5}, the photon-(heavy-)quark scattering contribution to
$\varphi$-dependent lepton-hadron deep-inelastic scattering (DIS) was investigated. It
turned out that, contrary to the basic photon-gluon fusion component, the
quark-scattering mechanism is practically $\cos(2\varphi)$-independent. This is due to
the fact that the quark-scattering contribution to the $\cos(2\varphi)$ asymmetry is, for
kinematic reasons, absent at LO and is negligibly small at NLO, of the order of $1\%$.
This indicates that the azimuthal distributions in charm leptoproduction could be a good
probe of the charm PDF in the proton.

In the present paper, we continue the studies of perturbatively stable observables by
considering the photon-gluon fusion mechanism in heavy-quark leptoproduction,
\begin{equation}
\ell(l )+N(p)\rightarrow \ell(l -q)+Q(p_{Q})+X[\bar{Q}](p_{X}). \label{1}
\end{equation}
In the case of unpolarized initial states and neglecting the contribution of $Z$-boson exchange,
the cross section of reaction (\ref{1}) can be written as
\begin{eqnarray}
\frac{\text{d}^{2}\sigma_{lN}}{\text{d}x\,\text{d}Q^{2}}&=&\frac{4\pi
\alpha^{2}_{\mathrm{em}}}{Q^{4}}\left\{ \left[ 1+(1-y)^{2}\right] F_{T}( x,Q^{2})
+2\left(1-y\right) F_{L}(x,Q^{2})\right\}  \nonumber \\
&=&\frac{2\pi \alpha^{2}_{\mathrm{em}}}{xQ^{4}}\left\{ \left[ 1+(1-y)^{2}\right] F_{2}( x,Q^{2})
-2xy^{2} F_{L}(x,Q^{2})\right\},  \label{2}
\end{eqnarray}
where $\alpha_{\mathrm{em}}$ is Sommerfeld's fine-structure constant,
$F_{2}(x,Q^2)=2x(F_{T}+F_{L})$ and the kinematic variables are defined by
\begin{eqnarray}
\bar{S}=\left( \ell +p\right) ^{2},\qquad &Q^{2}=-q^{2},\qquad &x=\frac{Q^{2}}
{2p\cdot q},  \nonumber \\
y=\frac{p\cdot q}{p\cdot \ell },\qquad \quad ~ &Q^{2}=xy\bar{S},\qquad &\xi=
\frac{Q^2}{m^2}.  \label{3}
\end{eqnarray}
In this paper, we investigate radiative corrections to the Callan-Gross ratio
in heavy-quark leptoproduction, defined as
\begin{equation}
R(x,Q^{2})=\frac{F_{L}(x,Q^{2})}{F_{T}(x,Q^{2})}. \label{4}
\end{equation}
First, we consider the exact NLO corrections to the quantity $R(x,Q^2)$ at low and
moderate $Q^2\lesssim m^2$ using explicit results \cite{LRSN,RSN}. Then, we analyze the
high-$Q^2$ regime with the help of the asymptotic NLO predictions for the structure
functions $F_T(x,Q^2)$ and $F_L(x,Q^2)$ presented in Refs.~\cite{BMSMN,BMSN}. Finally,
the soft-gluon (or threshold) contributions are investigated in the large-$x$ region in
the framework of the formalism developed in Ref.~\cite{Laenen-Moch}. To next-to-leading
logarithmic (NLL) accuracy, we calculate the NLO and NNLO soft-gluon corrections to both
structure functions. Our main results can be formulated as follows:
\begin{itemize}
\item Exact NLO corrections to the ratio $R(x,Q^{2})$ do not exceed 10$\%$ in the energy
range $x>10^{-4}$ at low and moderate $Q^2\lesssim m^2$.
 \item At high $Q^2\gg m^2$, the asymptotic NLO corrections to $R(x,Q^{2})$ are less than
10$\%$ for $10^{-4}<x<10^{-1}$.
 \item  At the NLL level, the NLO and NNLO soft-gluon predictions for $R(x,Q^{2})$ affect
 the LO results by less than a few percent at low and moderate $Q^2$ and $x\gtrsim 10^{-2}$.
 \item  In all the cases mentioned above, the NLO predictions for $R(x,Q^2)$ are
sufficiently insensitive,
 to within ten percent, to standard uncertainties in the QCD input parameters
$\mu_{F}$, $\mu_{R}$ and $\Lambda_{\mathrm{QCD}}$, and in the gluon PDF $g(x,\mu_F)$.
\end{itemize}
We conclude that, in contrast to the production cross sections, the Callan-Gross ratio
 in heavy-quark leptoproduction is an observable quantitatively well defined in pQCD.
 Perturbative stability of the photon-gluon fusion results for $R(x,Q^{2})$ is
 mainly due to the cancellation of large radiative corrections to the structure functions
$F_T(x,Q^2)$ and $F_L(x,Q^2)$ in their ratio, especially in the
 large-$x$ region. Measurements of the
 quantity $R(x,Q^2)$ in charm and bottom leptoproduction should provide a good test of the
 conventional parton model based on pQCD.

Concerning the experimental aspects, perturbative stability of the QCD predictions for
$R(x,Q^{2})$ observed in our studies is very useful for the extraction of the structure
functions $F_2^c(x,Q^2)$ and $F_2^b(x,Q^2)$ from the data. Usually, it is the so-called
``reduced cross section", $\tilde{\sigma}(x,Q^{2})$, that can directly be measured in DIS
experiments:
\begin{eqnarray}
\tilde{\sigma}(x,Q^{2})=\frac{1}{1+(1-y)^2}\frac{xQ^4}{2\pi
\alpha^{2}_{\mathrm{em}}}\frac{\text{d}^{2}\sigma_{lN}}{\text{d}x\text{d}Q^{2}}&=&F_{2}(
x,Q^{2})-\frac{2xy^{2}}{1+(1-y)^2}F_{L}(x,Q^{2}) \label{5} \\
&=&F_{2}(x,Q^{2})\left[1-\frac{y^{2}}{1+(1-y)^2}R_{2}(x,Q^{2}) \right], \label{6}
\end{eqnarray}
where
\begin{equation}
R_{2}(x,Q^{2})=2x\frac{F_{L}(x,Q^2)}{F_{2}(x,Q^2)}=\frac{R(x,Q^{2})}{1+R(x,Q^{2})}. \label{7}
\end{equation}

In earlier HERA analyses of charm and bottom electroproduction \cite{H1old}, the
corresponding longitudinal structure functions were taken to be zero for simplicity. In
this case, $\tilde{\sigma}(x,Q^{2})=F_{2}(x,Q^{2})$. In recent papers
\cite{H1HERA1,H1HERA2}, the structure function $F_{2}(x,Q^2)$ is evaluated from the
reduced cross section (\ref{5}) where the longitudinal structure function $F_{L}(x,Q^2)$
is estimated from the NLO QCD expectations. Instead of this rather cumbersome procedure,
we propose to use the expression (\ref{6}) with the quantity $R_{2}(x,Q^2)$ calculated in
LO approximation. This simplifies the extraction of $F_{2}(x,Q^2)$ from measurements of 
$\tilde{\sigma}(x,Q^{2})$ but does not affect the accuracy of the result in practice.

Indeed, the LO corrections to the extracted function $F_{2}(x,Q^2)$ due to the non-zero
value of $R_{2}(x,Q^2)$ cannot exceed 30$\%$ because the ratio $R_{2}(x,Q^{2})$ is itself
less than 0.3 practically in the entire region of the variables $x$ and $Q^2$. For this
reason, the NLO corrections to $R_{2}(x,Q^2)$, having a relative size of the order of
10\%, cannot affect the value of $F_{2}(x,Q^2)$ by more than 3\%. In reality, the effect
of radiative corrections to $R_{2}(x,Q^2)$ on the extracted values of $F_{2}(x,Q^2)$ is
less than 1$\%$ since $y\ll 1$ in most of the experimentally accessible kinematic range.

In the present paper, we derive compact hadron-level LO predictions for the ratio
$R_{2}(x,Q^{2})$ in the limit of low $x\to 0$. Assuming the low-$x$ asymptotic behavior
of the gluon PDF to be of the type $g(x,Q^2)\propto 1/x^{1+\delta}$, we provide analytic 
result for the ratio $R_{2}(x\to 0,Q^{2})\equiv R^{(\delta)}_2(Q^2)$ for arbitrary values 
of the parameter $\delta$ in terms of the Gauss hypergeometric function. 
Furthermore, we consider compact formulae for
$R^{(\delta)}_2(Q^2)$ in two particular cases: $\delta=1/2$ and $\delta=0$. The simplest case, 
$\delta =0$, which has already been studied recently in Ref.~\cite{kotikov}, leads to a 
non-singular behavior of the structure functions for $x\to 0$. The second choice, $\delta=1/2$, 
historically originates from the BFKL resummation of the leading powers of $\ln(1/x)$ \cite{BFKL1}.

In principle, the parameter $\delta$ is a function of $Q^2$ and this dependence is calculated 
using the DGLAP evolution equations \cite{DGLAP}. However, our analysis shows that hadron-level 
predictions for 
$R_2(x\to 0,Q^2)$ depend weakly on $\delta$ practically in the entire region of $Q^2$ 
for $\delta > 0.2$. In particular, the relative difference between $R^{(0.5)}_2(Q^2)$ and 
$R^{(0.3)}_2(Q^2)$ is less than few percent at $Q^2\gtrsim m^2$.
For this reason, our simple formula for $R^{(\delta)}_2(Q^2)$ with $\delta =1/2$ 
(i.e., without any evolution) describes with good accuracy the low-$x$ predictions for 
$R_2(x,Q^2)$ of the CTEQ PDF versions \cite{CTEQ6,CTEQ5}. We see that the hadron-level 
predictions for $R_2(x\to 0,Q^2)$ are stable not only under the NLO corrections to 
the partonic cross sections, but also under the DGLAP evolution of the gluon PDF. 

Finally, we show that our compact LO formulae for $R^{(\delta)}_2(Q^2)$ conveniently reproduce the 
HERA results for $F_2^c(x,Q^2)$ and $F_2^b(x,Q^2)$ obtained by H1 Collaboration \cite{H1HERA1,H1HERA2}
with the help of more cumbersome NLO estimations of $F_{L}(x,Q^2)$.

This paper is organized as follows. In Section~\ref{NLO}, we analyze the exact NLO results
for the Callan-Gross ratio at low and moderate $Q^2\lesssim m^2$ and the asymptotic NLO
predictions at high $Q^2\gg m^2$. The soft-gluon contributions to $R(x,Q^{2})$ are
investigated in Section~\ref{SGR}. To NLL accuracy, we calculate the threshold
NLO and NNLO corrections to both structure functions $F_T(x,Q^2)$ and $F_L(x,Q^2)$. The analytic LO
results for the ratio $R_{2}(x,Q^{2})$ at low $x$ are discussed in
Section~\ref{analytic}. 

\boldmath
\section{\label{NLO} NLO Predictions for the Callan-Gross Ratio $R(x,Q^2)$}
\unboldmath

\subsection{Born-Level Cross Sections}

\begin{figure}
\begin{center}
\mbox{\epsfig{file=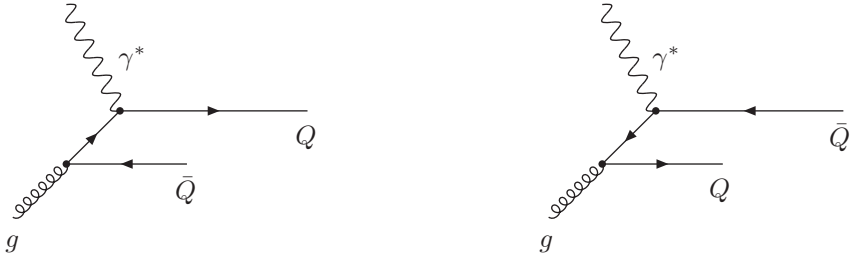,width=355pt}}
\end{center}
\caption{\label{Fg.1}\small Feynman diagrams of photon-gluon fusion at LO.}
\end{figure}
At LO, ${\cal O}(\alpha_{\mathrm{em}}\alpha_{s})$, leptoproduction of heavy flavors
proceeds through the photon-gluon fusion mechanism,
\begin{equation} \label{8}
\gamma ^{*}(q)+g(k_{g})\rightarrow Q(p_{Q})+\bar{Q}(p_{\bar{Q}}).
\end{equation}
The relevant Feynman diagrams are depicted in Fig.~\ref{Fg.1}.
The $\gamma ^{*}g$ cross sections, $\hat{\sigma}_k^{(0)}(z,\lambda)$ ($k=2,T,L$), have the
form \cite{LW1}:
\begin{eqnarray}
\hat{\sigma}_{2}^{(0)}(z,\lambda)&=&\frac{\alpha_{s}(\mu_R^2)}{2\pi}\hat{\sigma}_{B}(z)
\left\{\left[(1-z)^{2}+z^{2}+4\lambda z(1-3z)-8\lambda^{2}z^{2}\right]
\ln\frac{1+\beta_{z}}{1-\beta_{z}}-
\left[1+4z(1-z)(\lambda-2)\right]\beta_{z}\right\},  \label{9}\\
\hat{\sigma}_{L}^{(0)}(z,\lambda)&=&\frac{2\alpha_{s}(\mu_R^2)}{\pi}\hat{\sigma}_{B}(z)z
\left\{-2\lambda z\ln\frac{1+\beta_{z}}{1-\beta_{z}}+\left(1-z\right)\beta_{z}\right\},
\label{10}\\
\hat{\sigma}_{T}^{(0)}(z,\lambda)&=&\hat{\sigma}_{2}^{(0)}(z,\lambda)
-\hat{\sigma}_{L}^{(0)}(z,\lambda),
\end{eqnarray}
with
\begin{equation}\label{11}
\hat{\sigma}_{B}(z)=\frac{(2\pi)^2e_{Q}^{2}\alpha_{\mathrm{em}}}{Q^{2}}z,
\end{equation}
where $e_{Q}$ is the electric charge of quark $Q$ in units of the positron charge and
$\alpha_{s}(\mu_R^2)$ is the strong-coupling constant.
In Eqs.~(\ref{9})--(\ref{11}), we use the following definition of partonic kinematic
variables:
\begin{equation}\label{12}
z=\frac{Q^{2}}{2q\cdot k_{g}},\qquad\lambda =\frac{m^{2}}{Q^{2}}, \qquad
\beta_{z}=\sqrt{1-\frac{4\lambda z}{1-z}}.
\end{equation}
The hadron-level cross sections, $\sigma_{k}(x,Q^2)$ ($k=2,T,L$), have the form
\begin{equation}
\sigma_{k}(x,Q^2)=\int\limits_{x(1+4\lambda)}^{1}\text{d}z\,g(z,\mu_{F})
\hat{\sigma}_{k}\left(\frac{x}{z},\lambda,\mu_{F},\mu_{R}\right),
\label{13}
\end{equation}
where $g(z,\mu_{F})$ is the gluon PDF of the proton. The leptoproduction cross sections
$\sigma_{k}(x,Q^2)$ are related to the structure functions $F_{k}(x,Q^2)$ as follows:
\begin{eqnarray}
F_{k}(x,Q^2) &=&\frac{Q^{2}}{8\pi^{2}\alpha_{\mathrm{em}}x}\sigma_{k}(x,Q^2)
\qquad (k=T,L), \\
F_{2}(x,Q^2) &=&\frac{Q^{2}}{4\pi^{2}\alpha_{\mathrm{em}}}\sigma_{2}(x,Q^2).
\label{15}
\end{eqnarray}

\boldmath
\subsection{Exact NLO Predictions at Low and Moderate $Q^2$}
\unboldmath

At NLO, ${\cal O}(\alpha_{\mathrm{em}}\alpha_{s}^2)$, the contribution of the photon-gluon
component is usually presented in terms of the dimensionless coefficient functions
$c_{k}^{(n,l)}(z,\lambda)$ ($k=T,L$), as
\begin{equation}\label{16}
\hat{\sigma}_{k}(z,\lambda,m^2,\mu^{2})=\frac{e_{Q}^{2}\alpha_{\mathrm{em}}\alpha_{s}(\mu
^{2})}{m^{2}}\left\{ c_{k}^{(0,0)}(z,\lambda)+ 4\pi\alpha_{s}(\mu^{2})\left[
c_{k}^{(1,0)}(z,\lambda)+c_{k}^{(1,1)}(z,\lambda)\ln\frac{\mu^{2}}{m^{2}}
\right]\right\}+{\cal O}(\alpha_{s}^2).
\end{equation}
where we identify $\mu=\mu_{F}=\mu_{R}$.

In this paper, we neglect the $\gamma ^{*}q(\bar{q})$ fusion subprocesses. This is
justified as their contributions to heavy-quark leptoproduction vanish at LO and are
small at NLO \cite{LRSN}. To be precise, the light-quark-initiated corrections to both 
$F_T$ and $F_L$ structure functions are negative and less than $10\%$ in a wide 
kinematic range \cite{LRSN}. 
Our estimates show that these contributions cancel in the ratio $R(x,Q^2)=F_L/F_T$ 
with an accuracy less than few percent.

The coefficients $c_{T}^{(1,1)}(z,\lambda)$ and $c_{L}^{(1,1)}(z,\lambda)$ of the
$\mu$-dependent logarithms can be evaluated explicitly using renormalization group
arguments \cite{LRSN,Laenen-Moch}. The results of direct calculations of the coefficient
functions $c_{k}^{(1,0)}(z,\lambda)$ ($k=T,L$) are presented in Refs.~\cite{LRSN,RSN}.
Using these NLO predictions, we compute the $x$ dependence of the ratio
$R(x,Q^2)=F_L/F_T$ at several values of $\xi=1/\lambda=Q^2/m^2$.

\begin{figure}
\begin{center}
\begin{tabular}{cc}
\mbox{\epsfig{file=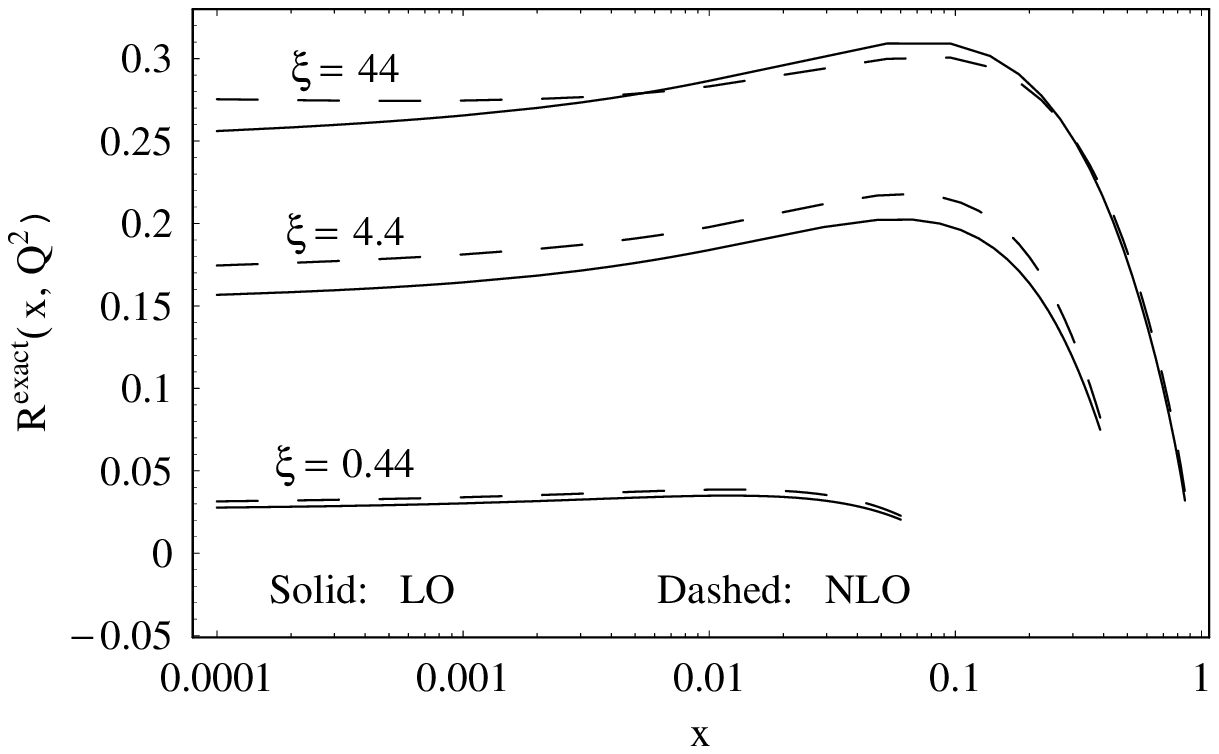,width=250pt}}
& \mbox{\epsfig{file=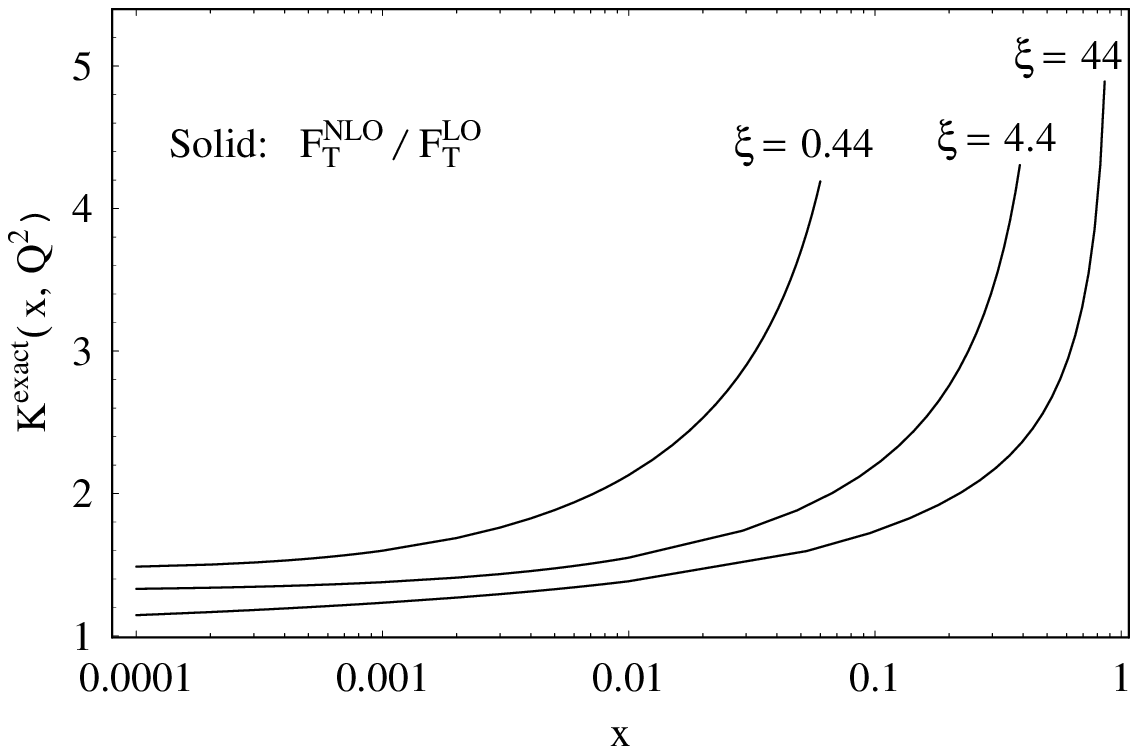,width=250pt}}\\
\end{tabular}
\caption{\label{Fg.2}\small \emph{Left panel:} $x$ dependence of the Callan-Gross
ratio, $R(x,Q^2)=F_L/F_T$, in charm leptoproduction for $\xi =0.44$, 4.4 and 44
at LO (solid lines) and NLO (dashed lines). \emph{Right panel:} $x$ dependence of the $K$
factor for the transverse structure function,
$K(x,Q^2)=F_T^{\mathrm{NLO}}/F_T^{\mathrm{LO}}$, at the same values of $\xi$.}
\end{center}
\end{figure}
The left panel of Fig.~\ref{Fg.2} shows the Callan-Gross ratio $R(x,Q^2)$ as a function of
$x$ for $\xi =0.44$, 4.4 and 44.
In our calculations, we use the CTEQ5M parametrization of the gluon PDF together with the
values $m_c=1.3$~GeV and $\Lambda_{3}=373$~MeV \cite{CTEQ5}.
Unless otherwise stated, we use $\mu=\sqrt{4m_c^{2}+Q^{2}}$ throughout this paper.

For comparison, the right panel of Fig.~\ref{Fg.2} shows the $x$ dependence of the QCD
correction factor for the transverse structure function,
$K(x,Q^2)=F_T^{\mathrm{NLO}}/F_T^{\mathrm{LO}}$. One can see that large
radiative corrections to the structure functions $F_T(x,Q^2)$ and $F_L(x,Q^2)$,
especially at non-small $x$,
cancel each
other in their ratio $R(x,Q^2)=F_L/F_T$ with good accuracy. As a result, the NLO
contributions to the ratio $R(x,Q^2)$ are less than $10\%$ for $x\gtrsim 10^{-4}$ at low
and moderate $Q^2\lesssim m_{c}^{2}$.

Another remarkable property of the Callan-Gross ratio closely related to fast
perturbative convergence is its parametric stability.\footnote{Of course, parametric
stability of the fixed-order results does not imply a fast convergence of the
corresponding series. However, a fast convergent series must be parametrically stable. In
particular, it must exhibit feeble $\mu_{F}$ and $\mu_{R}$ dependences.}
Our analysis shows that
the fixed-order predictions for the ratio $R(x,Q^2)$ are less sensitive to standard
uncertainties in the QCD input parameters than the corresponding ones for the production
cross sections. For instance, sufficiently above the production threshold, changes of
$\mu$ in the range $(1/2)\sqrt{4m_{c}^{2}+Q^{2}}<\mu <2 \sqrt{4m_{c}^{2}+Q^{2}}$
only lead to $10\%$ variations of $R(x,Q^{2})$ at NLO. For comparison, at
$x=0.1$ and $\xi = 4.4$, such changes of $\mu$ affect the NLO predictions for the
quantities $F_{T}(x,Q^2)$ and $R(x,Q^{2})$ in charm leptoproduction by more than $100\%$
and less than $10\%$, respectively.

Keeping the value of the variable $Q^{2}$ fixed, we analyze the dependence of the pQCD
predictions on the uncertainties in the heavy-quark mass. Sufficiently above the production
threshold, i.e.\ in the plateau regions of the $x$ distributions of $R(x,Q^{2})$ in
Fig.~\ref{Fg.2}, changes of the charm-quark mass in the interval 1.3~GeV${}<m_{c}<1.7$~GeV
affect the Callan-Gross ratio by 2\%--3\% at $Q^{2}=10$~GeV$^2$. The corresponding
variations of the structure functions $F_T(x,Q^2)$ and $F_L(x,Q^2)$ are about 20\%.
We also
verify that all the CTEQ versions \cite{CTEQ6,CTEQ5} of the gluon PDF lead to
NLO predictions for $R(x,Q^{2})$ that coincide with each other with an accuracy of about
$5\%$ practically at all $x\gtrsim 10^{-4}$.

\boldmath
\subsection{Asymptotic NLO Results at High $Q^2\gg m^2$}
\unboldmath

The analytic form of the heavy-quark coefficient functions for
lepton-hadron DIS in the kinematical regime $Q^2\gg m^2$ is presented in
Refs.~\cite{BMSMN,BMSN}. The calculations were performed up to NLO
in $\alpha_s$ using operator product expansion techniques.\footnote{For the longitudinal 
cross section $\hat{\sigma}_{L}(z,Q^2, m^2,\mu^{2})$, the asymptotic heavy-quark 
coefficient functions $a_{L}^{l,(n,m)}(z)$ are known up to NNLO in $\alpha_s$ 
\cite{Blumlein}.}
In the asymptotic regime $\xi \to \infty$, the production cross sections have the
following decomposition in terms of the coefficient functions $a_{k}^{l,(n,m)}(z)$
($k=2,L$):
\begin{equation}\label{18}
\hat{\sigma}_{k}(z,Q^2, m^2,\mu^{2})=\frac{e_{Q}^{2}\alpha_{\mathrm{em}}}{4\pi
m^{2}}\sum_{l=1}^{\infty }\left[ 4\pi \alpha_{s}(\mu ^{2})\right]
^{l}\sum_{m+n<l}^{n}a_{k}^{l,(n,m)}(z)\ln^{n}\frac{\mu
^{2}}{m^{2}}\ln^{m}\frac{Q^{2}}{m^{2}}+{\cal O}
\left(\frac{m^{2}}{Q^{2}}\right).
\end{equation}

It was found in Refs.~\cite{BMSMN,BMSN} that the hadron-level structure function
$F^{\text{asymp}}_{2}(x,Q^2)$ approaches, to within ten percent, the corresponding exact
value $F^{\text{exact}}_{2}(x,Q^2)$ for $\xi \gtrsim 10$ and $x<10^{-1}$ both at LO and NLO.
In the case of the longitudinal structure function $F^{\text{asymp}}_{L}(x,Q^2)$, the
approach to $F^{\text{exact}}_{L}(x,Q^2)$ starts at much larger values of $\xi \gtrsim
4\times 10^{2}$.

\begin{figure}
\begin{center}
\begin{tabular}{cc}
\mbox{\epsfig{file=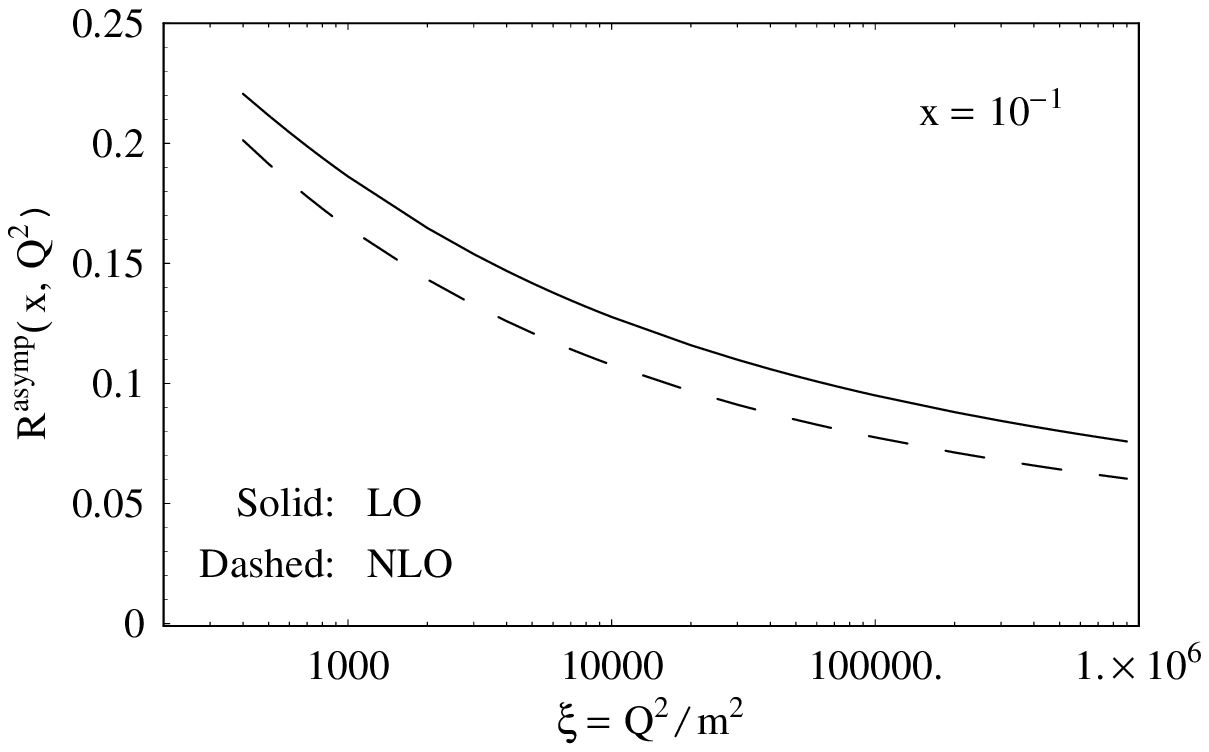,width=250pt}}
& \mbox{\epsfig{file=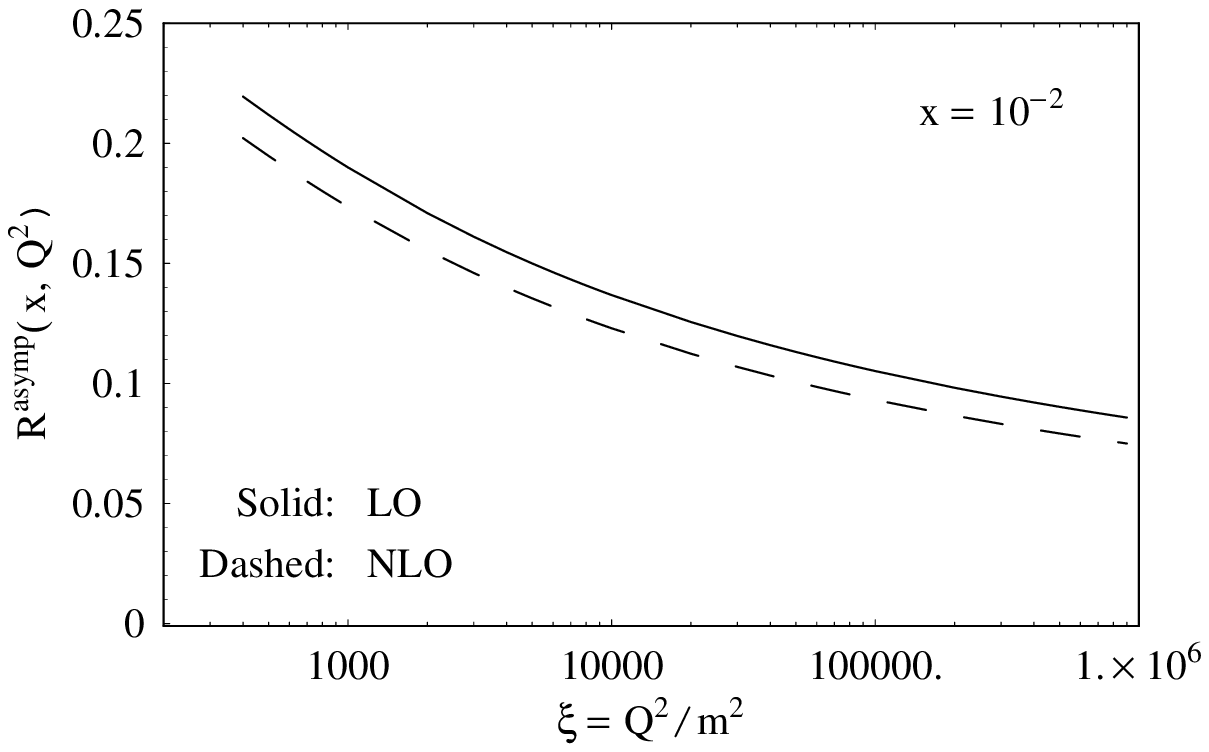,width=250pt}}\\
\mbox{\epsfig{file=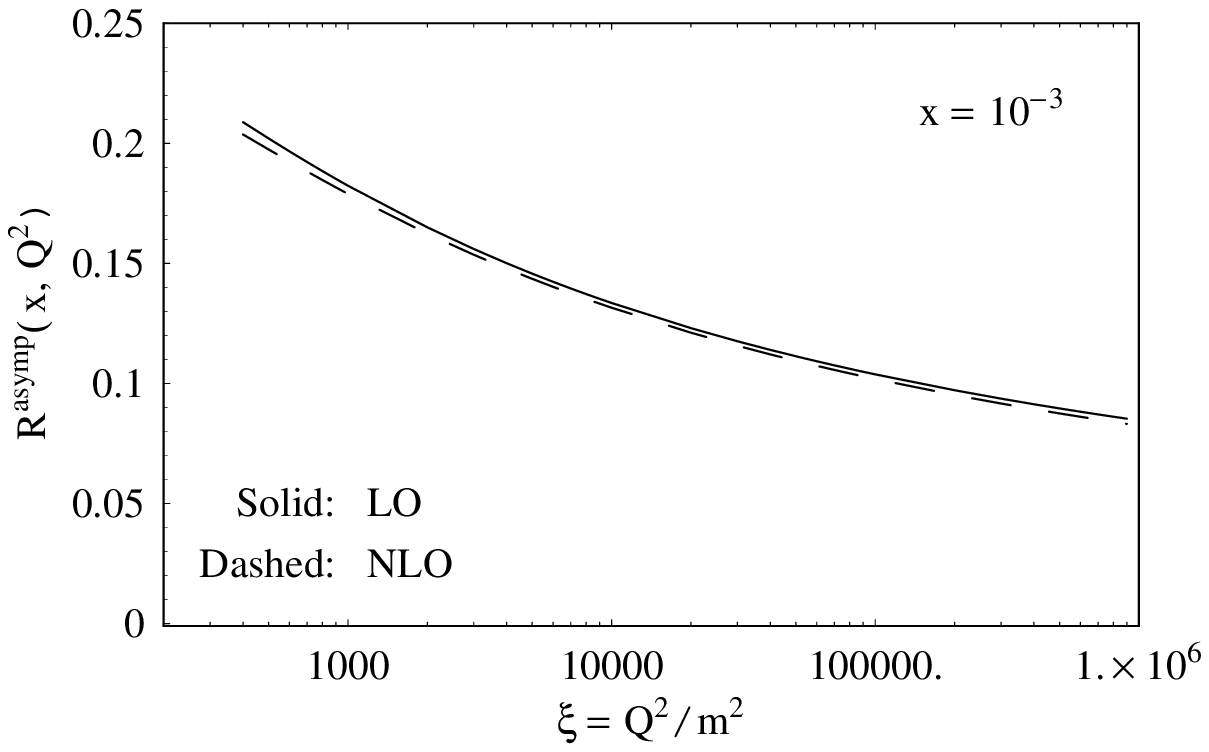,width=250pt}}
& \mbox{\epsfig{file=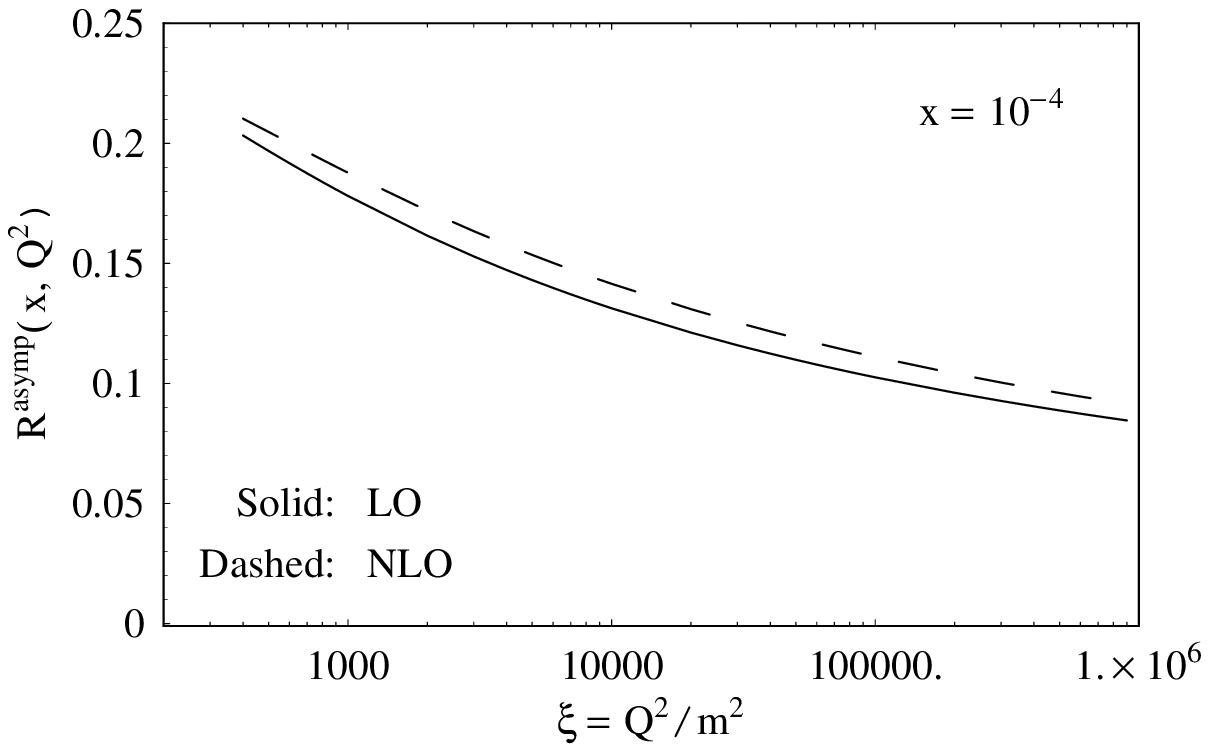,width=250pt}}\\
\end{tabular}
\caption{\label{Fg.3}\small $Q^2$ dependence of the asymptotic high-$Q^2$ ($Q^2\gg m^2$)
predictions for the Callan-Gross ratio, $R(x,Q^2)=F_L/F_T$, in charm leptoproduction at
$x=10^{-1}$, $10^{-2}$, $10^{-3}$ and $10^{-4}$ in LO (solid lines) and NLO (dashed
lines).}
\end{center}
\end{figure}
\begin{figure}
\begin{center}
\mbox{\epsfig{file=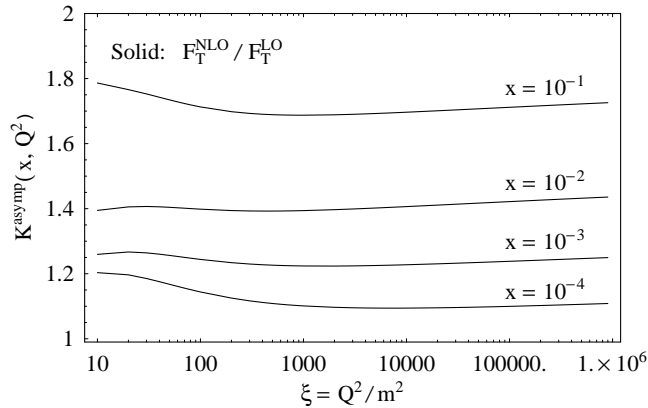,width=250pt}}
 \caption{\label{Fg.4}\small $Q^2$-dependence of the asymptotic high-$Q^2$ ($Q^2\gg m^2$)
predictions for the $K$ factor, $K(x,Q^2)=F_T^{\mathrm{NLO}}/F_T^{\mathrm{LO}}$, at
$x=10^{-1}$, $10^{-2}$, $10^{-3}$ and $10^{-4}$.}
\end{center}
\end{figure}

Using the analytic NLO results for the coefficient functions presented in
Ref.~\cite{BMSMN},
we calculate the asymptotic high-$Q^2$ behavior of the ratio $R(x,Q^2)=F_L/F_T$ at
several values of $x$. Figure~\ref{Fg.3} shows $R^{\text{asymp}}(x,Q^2)$ in charm
leptoproduction as a function of $\xi$ for $x=10^{-1}$, $10^{-2}$, $10^{-3}$ and
$10^{-4}$.
In Fig.~\ref{Fg.4}, we show the $Q^2$ dependence of the asymptotic predictions for the
$K$ factor $K(x,Q^2)=F_T^{\mathrm{NLO}}/F_T^{\mathrm{LO}}$ at the same values of $x$.
One can see that
the quantity $K(x,Q^2)$ is practically independent of $Q^2$ at fixed values of $x$ and
tends to unity at low $x$. This implies that perturbative stability of the Callan-Gross
ratio at low $x$ is due to the smallness of the radiative corrections to both structure
functions. At non-small $x$, the radiative corrections to $F_T(x,Q^2)$ and $F_L(x,Q^2)$
are large but cancel each other in their ratio $R(x,Q^2)=F_L/F_T$ with good accuracy.

\section{\label{SGR} Soft-Gluon Corrections at NLO and NNLO}

In this Section, we consider the NLO and NNLO predictions for the Callan-Gross ratio due
to the contribution of the photon-gluon fusion mechanism in the soft-gluon approximation
and propose an improvement. For the reader's convenience, we collect the final results
for the parton-level cross sections to NLL accuracy. More details may be found in
Refs.~\cite{Laenen-Moch,we2,we4}.

At NLO, photon-gluon fusion receives contributions from the virtual ${\cal
O}(\alpha_{\mathrm{em}}\alpha_{s}^{2})$ corrections to the Born process~(\ref{8}) and
from real-gluon emission,
\begin{equation}  \label{19}
\gamma ^{*}(q)+g(k_{g})\rightarrow Q(p_{Q})+\bar{Q}(p_{\bar{Q}})+g(p_{g}).
\end{equation}
The partonic invariants describing the single-particle inclusive (1PI) kinematics are
\begin{eqnarray}
s^{\prime }&=&2q\cdot k_{g}=s+Q^{2}=\zeta S^{\prime },\qquad \qquad t_{1}=\left(
k_{g}-p_{Q}\right) ^{2}-m^{2}=\zeta T_{1},  \nonumber \\
s_{4}&=&s^{\prime }+t_{1}+u_{1},\qquad \qquad \qquad \qquad ~~ u_{1}=\left(
q-p_{Q}\right) ^{2}-m^{2}=U_{1}, \label{20}
\end{eqnarray}
where $\zeta$ is defined through $\vec{k}_{g}= \zeta\vec{p}\,$ and $s_{4}$ measures the
inelasticity of the reaction (\ref{19}). The corresponding 1PI hadron-level variables
describing the reaction (\ref{1}) are
\begin{eqnarray}
S^{\prime }&=&2q\cdot p=S+Q^{2},\qquad \qquad T_{1}=\left( p-p_{Q}\right)
^{2}-m^{2},  \nonumber \\
S_{4}&=&S^{\prime }+T_{1}+U_{1},\qquad \qquad \quad U_{1}=\left( q-p_{Q}\right)
^{2}-m^{2}. \label{21}
\end{eqnarray}

The exact NLO calculations of unpolarized heavy-quark production in $\gamma g$
\cite{Ellis-Nason}, $\gamma ^{*}g$ \cite{LRSN}, and $gg$
\cite{Nason-D-E-1} collisions show that, near the partonic
threshold, a strong logarithmic enhancement of the cross sections takes place in the
collinear, $|\vec{p}_{g,T}|\to 0$, and soft, $|\vec{p}_{g}|\to 0$,
limits. This threshold (or soft-gluon) enhancement is of universal nature in perturbation
theory and originates from an incomplete cancellation of the soft and collinear
singularities between the loop and the bremsstrahlung contributions. Large leading and
next-to-leading threshold logarithms can be resummed to all orders of the perturbative
expansion using the appropriate evolution equations
\cite{Contopanagos-L-S}. The analytic results for the resummed
cross sections are ill-defined due to the Landau pole in the coupling constant
$\alpha_{s}$. However, if one considers the obtained expressions as generating functionals
and re-expands them at fixed order in $\alpha_{s}$, no
divergences associated with the Landau pole are encountered.

Soft-gluon resummation for the photon-gluon fusion was performed in
Ref.~\cite{Laenen-Moch} and confirmed in Refs.~\cite{we2,we4}. To NLL accuracy, the
perturbative expansion for the partonic cross sections,
$\mathrm{d}^{2}\hat{\sigma}_{k}(s^{\prime},t_{1},u_{1})/(\mathrm{d}t_{1}\,
\mathrm{d}u_{1})$ ($k=T,L$), can be written in factorized form as
\begin{equation}  \label{22}
s^{\prime 2}\frac{\text{d}^{2}\hat{\sigma}_{k}}{\text{d}t_{1}\text{d}u_{1}}( s^{\prime
},t_{1},u_{1}) =B_{k}^{\text{{\rm Born}}}( s^{\prime },t_{1},u_{1})\left[\delta
(s^{\prime }+t_{1}+u_{1}) +\sum_{n=1}^{\infty }\left( \frac{\alpha
_{s}C_{A}}{\pi}\right)^{n}K^{(n)}( s^{\prime },t_{1},u_{1})\right] ,
\end{equation}
with the Born-level distributions $B_{k}^{\text{{\rm Born}}}(s^{\prime },t_{1},u_{1})$
given by
\begin{eqnarray}
B_{T}^{\text{{\rm Born}}}( s^{\prime },t_{1},u_{1}) &=&\pi e_{Q}^{2}\alpha_{\mathrm{em}}\alpha
_{s}\left\{ \frac{t_{1}}{u_{1}}+\frac{u_{1}}{t_{1} }+4\left( \frac{s}{s^{\prime
}}-\frac{m^{2}s^{\prime }}{t_{1}u_{1}}\right) \left[ \frac{s^{\prime
}(m^{2}-Q^{2}/2)}{t_{1}u_{1}}+\frac{Q^{2}}{s^{\prime}
}\right] \right\} ,  \label{23} \\
B_{L}^{\text{{\rm Born}}}( s^{\prime },t_{1},u_{1}) &=&\pi e_{Q}^{2}\alpha_{\mathrm{em}}\alpha
_{s} \frac{8Q^{2}}{s^{\prime }}\left( \frac{s}{s^{\prime }}-\frac{m^{2}s^{\prime
}}{t_{1}u_{1}}\right) . \label{23a}
\end{eqnarray}
Note that the functions $K^{(n)}( s^{\prime },t_{1},u_{1}) $ in Eq.~(\ref{22}) originate
from the collinear and soft limits and are the same for both cross sections
$\hat{\sigma}_{T}$ and $\hat{\sigma}_{L}$. At NLO and NNLO, the soft-gluon corrections to
NLL accuracy in the $\overline{\text{MS}}$ scheme read
\begin{eqnarray}
K^{(1)}( s^{\prime },t_{1},u_{1}) &=&2\left[ \frac{\ln \left( s_{4}/m^{2}\right)
}{s_{4}}\right]_{+}-\left[ \frac{1}{s_{4}}\right]_{+}\left[ 1+\ln
\frac{u_{1}}{t_{1}} -\left( 1-\frac{2C_{F}}{ C_{A}}\right) \left(
1+\text{Re}L_{\beta }\right) +\ln \frac{\mu ^{2}
}{m^{2}} \right]   \nonumber \\
&&{}+\delta ( s_{4}) \ln \frac{-u_{1}}{m^{2}} \ln \frac{\mu ^{2}}{m^{2}}, \label{24} \\
K^{(2)}\left( s^{\prime} ,t_{1},u_{1}\right)  &=&2\left[ \frac{\ln ^{3}\left(
s_{4}/m^{2}\right) }{s_{4}}\right]_{+}  \nonumber \\
&&{}-3\left[ \frac{\ln ^{2}\left( s_{4}/m^{2}\right) }{s_{4}}\right]_{+}\left[ 1+\ln
\frac{u_{1}}{t_{1}} -\left( 1-\frac{2C_{F}}{ C_{A}}\right) \left(
1+\text{Re}L_{\beta }\right) +\frac{2}{3}\frac{b_{2}}{
C_{A}}+\ln\frac{\mu ^{2}}{m^{2}} \right]  \nonumber \\
&&{}+2\left[ \frac{\ln \left( s_{4}/m^{2}\right) }{s_{4}}\right]_{+} \left[ 1+ \ln
\frac{u_{1}}{t_{1}} -\left( 1-\frac{2C_{F}}{C_{A}}\right) \left(
1+\text{Re}L_{\beta }\right) +\ln \frac{-u_{1}}{m^{2}}
+\frac{b_{2}}{C_{A}}+\frac{1
}{2}\ln\frac{\mu ^{2}}{m^{2}} \right]   \nonumber \\
&&{}\times\ln \frac{\mu ^{2}}{m^{2}}-\left[ \frac{1}{s_{4}}\right]_{+} \ln
^{2}\frac{\mu ^{2}}{m^{2}} \left[ \ln\frac{-u_{1}}{m^{2}}
+\frac{b_{2}}{2 C_{A}}\right] ,  \label{25}
\end{eqnarray}
where $b_{2}=\left( 11C_{A}-2n_{f}\right) /12$ is the first coefficient of the beta
function,
\begin{equation}
\beta \left( \alpha_{s}\right) = \frac{\text{d}\ln \alpha_{s}\left( \mu
^{2}\right) }{\text{d}\ln \mu ^{2}}=-\sum_{k=1}^{\infty }b_{k+1}
\left(\frac{\alpha_{s}}{\pi}\right)^{k}.
\label{26}
\end{equation}
In Eqs.~(\ref{24}) and
(\ref{25}), $C_{A}=N_{c}$, $ C_{F}=(N_{c}^{2}-1)/(2N_{c})$, $n_{f}$ is the number of
active quark flavors, $N_{c}$ is the number of quark colors, and $ L_{\beta }=(1-2m^{2}/s)\{\ln
[(1-\beta_{z})/(1+\beta_{z})]+$i$\pi\}$ with $\beta_{z}=\sqrt{1-4m^{2}/s}$. The
single-particle inclusive ``plus'' distributions are defined by
\begin{equation}  \label{27}
\left[\frac{\ln^{l}\left( s_{4}/m^{2}\right) }{s_{4}}\right]_{+}=\lim_{\epsilon
\rightarrow 0}\left[\frac{\ln^{l}\left(s_{4}/m^{2}\right) }{s_{4}}\theta (
s_{4}-\epsilon)+\frac{1}{l+1}\ln ^{l+1}\frac{\epsilon }{m^{2}} \delta (
s_{4})\right].
\end{equation}
For any sufficiently regular test function $h(s_{4})$, Eq.~(\ref{27}) implies that
\begin{equation}\label{28}
\int\limits_{0}^{s_{4}^{\max }}\text{d}s_{4}\,h(s_{4})\left[ \frac{\ln ^{l}\left(
s_{4}/m^{2}\right) }{s_{4}}\right]_{+}=\int\limits_{0}^{s_{4}^{\max
}}\text{d}s_{4}\left[ h(s_{4})-h(0)\right] \frac{\ln ^{l}\left( s_{4}/m^{2}\right)
}{s_{4}}+\frac{1}{l+1}h(0)\ln ^{l+1}\frac{s_{4}^{\max }}{m^{2}} .
\end{equation}

In Eqs.~(\ref{24}) and (\ref{25}), we have also preserved the NLL terms for the
scale-dependent logarithms. Note that Eqs.~(\ref{23})--(\ref{24})
agree to NLL accuracy with the exact ${\cal O}(\alpha_{\mathrm{em}}\alpha_{s}^{2})$
calculations
of the photon-gluon cross sections $\hat{\sigma}_{T}$ and $\hat{\sigma}_{L}$ given in
Ref.~\cite{LRSN}.

Numerical investigation of the results (\ref{23})--(\ref{25}) was performed in
Refs.~\cite{Laenen-Moch,we4}. It was shown that soft-gluon corrections reproduce
satisfactorily the threshold behavior of the available exact results for the partonic
cross section $\hat{\sigma}_{2}=\hat{\sigma}_{T}+\hat{\sigma}_{L}$ at
$\xi\lesssim 1$. Since the gluon PDF supports just the
threshold region, the soft-gluon contribution dominates the hadron-level structure
function $F_{2}$ at energies not so far from the production threshold. It was shown in
Ref.~\cite{Laenen-Moch} that Eqs.~(\ref{23}) and (\ref{24}) render it possible to describe with
good accuracy the exact NLO predictions \cite{LRSN} for the function $F_{2}(x,Q^2)$ at
$x\gtrsim 10^{-3}$ and relatively low virtuality $Q^{2}\sim m^{2}$.

In the present paper, we analyze separately the partonic cross sections
$\hat{\sigma}_{T}$ and $\hat{\sigma}_{L}$. It turns out that the quality of the adopted
soft-gluon approximation is worse for $\hat{\sigma}_{L}$ than for $\hat{\sigma}_{T}$. To
clarify the situation, let us remember that the NLL approximation allows us to determine
unambiguously only the singular $s_{4}$ behavior of the cross sections defined by
Eq.~(\ref{27}). This implies that the $s_{4}$ dependence of the Born-level distributions
$\left.B_{T,L}^{\text{{\rm Born}}}(
s^{\prime},t_{1},u_{1})\right|_{u_{1}=s_{4}-s^{\prime}-t_{1}}$ is chosen quite
arbitrarily in Eqs.~(\ref{23}) and (\ref{23a}). To improve the situation, we propose the
following procedure to determine the $s_{4}$ dependence of the differential cross
sections based on a comparison of the soft-gluon predictions with the exact NLO results.
First, we define the on-shell Born-level distributions in the LO kinematics, i.e.\ at
$s_{4}=0$, as $\tilde{B}_{T,L}^{\text{{\rm Born}}}( s^{\prime},t_{1})= \left.
B_{T,L}^{\text{{\rm Born}}}( s^{\prime},t_{1},u_{1})\right|_{u_{1}=-s^{\prime}-t_{1}}$.
Then we introduce new quantities, $\hat{B}_{T,L}^{\text{{\rm
Born}}}(s^{\prime},t_{1},u_{1})$, with the following $s_{4}$ dependence:
$\hat{B}_{T,L}^{\text{{\rm
Born}}}(s^{\prime},t_{1},u_{1})\equiv\tilde{B}_{T,L}^{\text{{\rm
Born}}}(x_{4}s^{\prime},x_{4}t_{1})$, where
$x_{4}=-u_{1}/(s^{\prime}+t_{1})=1-s_{4}/(s^{\prime}+t_{1})$. Comparison with the exact
NLO results given by Eqs.~(4.7) and (4.8) in Ref.~\cite{LRSN} indicates that the usage of
the distributions $\hat{B}_{T,L}^{\text{{\rm Born}}}(s^{\prime},t_{1},u_{1})$ instead of
$B_{T,L}^{\text{{\rm Born}}}(s^{\prime},t_{1},u_{1})$ leads to a more accurate account of
the leading-logarithmic (LL) and NLL contributions originating from collinear gluon
emission. Our numerical analysis shows that the new quantities $\hat{B}_{T,L}^{\text{{\rm
Born}}}(s^{\prime},t_{1},u_{1})$ improve essentially the quality of the soft-gluon
approximation for both $\hat{\sigma}_{T}$ and $\hat{\sigma}_{L}$. More details can be
found in Ref.~\cite{we_prep}. In our further studies, we use the improved Born-level
distributions, $\hat{B}_{T,L}^{\text{{\rm Born}}}(s^{\prime},t_{1},u_{1})$, instead of
old ones given by Eqs.~(\ref{23}) and (\ref{23a}).

Note that the redefinition of the usual Born-level distributions used in the present 
paper does not affect any previous predictions of the standard resummation approach. The 
only purpose of our redefinition is to extend the region of applicability of the 
soft-gluon approximation to higher values of $Q^2$.

\begin{figure}
\begin{center}
\begin{tabular}{cc}
\mbox{\epsfig{file=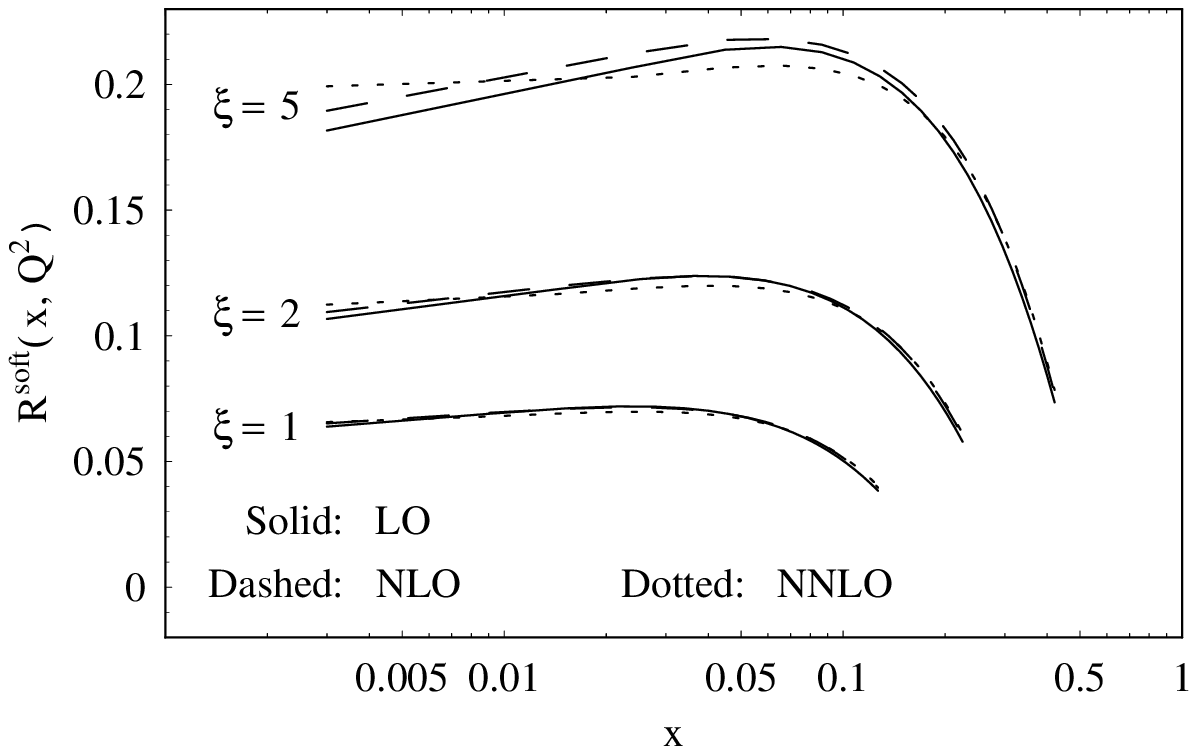,width=250pt}}
& \mbox{\epsfig{file=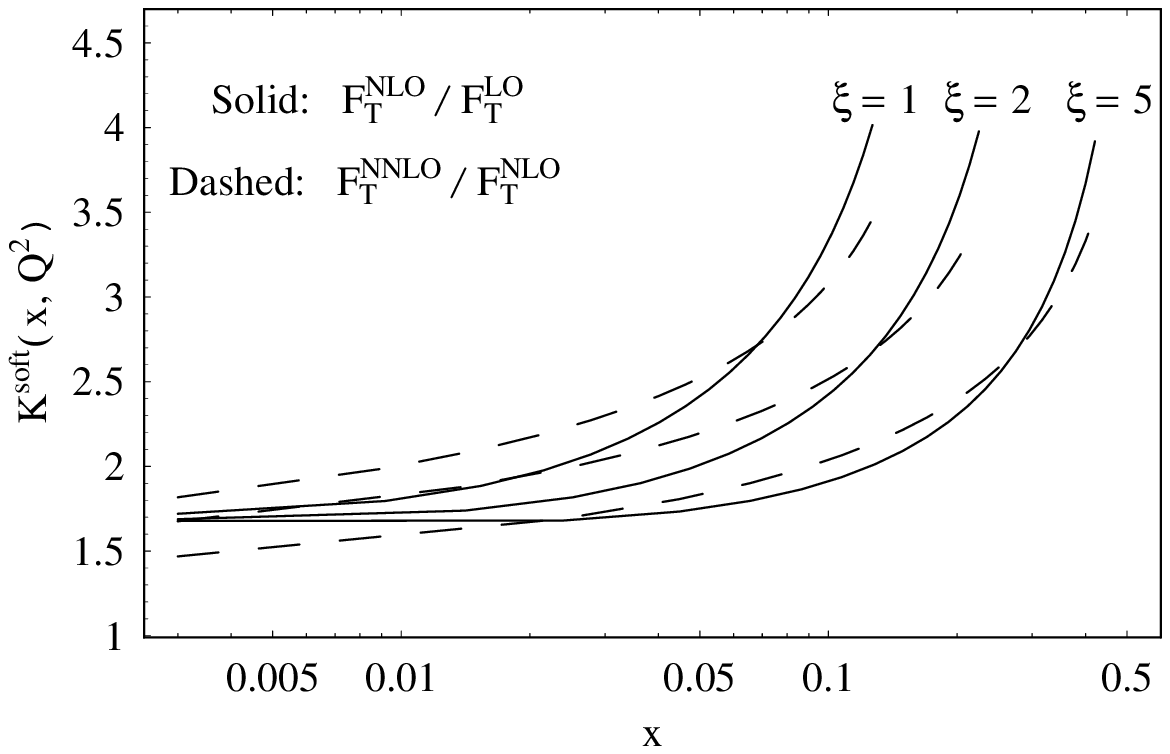,width=250pt}}\\
\end{tabular}
\caption{\label{Fg.5}\small \emph{Left panel:} LO (solid lines), NLO (dashed lines) and
NNLO (dotted lines) soft-gluon predictions for the $x$ dependence of the Callan-Gross ratio,
$R(x,Q^2)=F_L/F_T$, in charm leptoproduction at $\xi=1$, 2 and 5.
\emph{Right panel:} $x$ dependence of the $K$ factors
$K^{(1)}(x,Q^2)=F_T^{\mathrm{NLO}}/F_T^{\mathrm{LO}}$ (solid line) and
$K^{(2)}(x,Q^2)=F_T^{\mathrm{NNLO}}/F_T^{\mathrm{NLO}}$ (dashed curve) for the
transverse structure function at the same values of $\xi$.}
\end{center}
\end{figure}
Our results for the $x$ distributions of the Callan-Gross ratio $R(x,Q^2)=F_L/F_T$ in
charm leptoproduction are presented at several values of $\xi$ in the left panel of
Fig.~\ref{Fg.5}. For comparison, the $K$ factors
$K^{(1)}(x,Q^2)=F_T^{\mathrm{NLO}}/F_T^{\mathrm{LO}}$ and
$K^{(2)}(x,Q^2)=F_T^{\mathrm{NNLO}}/F_T^{\mathrm{NLO}}$ for the transverse structure
function are shown at the same values of $\xi$ in the right panel of Fig.~\ref{Fg.5}. One
can see that the sizeable soft-gluon corrections to the production cross sections affect
the Born predictions for $R(x,Q^2)$ both at NLO and NNLO very little, by a few percent
only.

Let us briefly discuss the origin of the perturbative stability of the Callan-Gross ratio. 
Note that the mere spin-independent structure of the Sudakov logarithms can not explain 
our results, since perturbative stability does not take place at the parton level. 
In fact, the ratios $\frac{c_{L}^{(1,0)}}{c_{T}^{(1,0)}}(z,Q^{2})$ and 
$\frac{c_{L}^{(0,0)}}{c_{T}^{(0,0)}}(z,Q^{2})$ differ essentially from each other, even near the 
threshold. This is due to the fact that, according to Eq.~(\ref{22}), the soft-gluon corrections 
are determined by convolutions of the Born cross sections with the Sudakov logarithms, 
which, apart from factorized $\delta(s_{4})$ terms, contain also nonfactorizable ones, see 
Eq.~(\ref{28}). For instance, values of $z \sim 10^{-1}$ allow $s_{4}/m^{2}\sim 1$ at $Q^{2}\sim m^{2}$,
which leads to significant nonfactorizable corrections. In other words, collinear bremsstrahlung 
carries away a large part of the initial energy. Since the longitudinal and transverse Born-level 
partonic cross sections have different energy behaviors, the so-called soft-gluon radiation has 
different impacts on these quantities.

Our analysis shows that two more factors are responsible for perturbative stability of the 
hadron-level ratio $R(x,Q^2)$. First, for relatively low virtuality $Q^{2}\sim m^{2}$,
both $\hat{\sigma}_{T}(z,Q^{2})$ and $\hat{\sigma}_{L}(z,Q^{2})$ take their maximum 
values practically at the same values of $z$ not far from the threshold. 
Second, at $x\sim 10^{-2}$--$10^{-1}$, the gluon distribution function supports just the
threshold region contribution. According to the saddle point arguments, both these factors
together lead to an approximate factorization of the Sudakov logarithms at the hadron level 
and essential cancellation of their contributions in the ratio $R(x,Q^2)=F_L/F_T$. 

Note also that the situation with perturbative stability of the Callan-Gross ratio is very 
similar to the corresponding one that takes place for the azimuthal asymmetry in heavy-quark 
photo- and leptoproduction. In detail, the soft-gluon corrections to the azimuthal asymmetry were 
considered in Refs.~\cite{we2,we4}.

\boldmath
\section{\label{analytic} Analytic LO Predictions at low $x$}
\unboldmath

Since the radiative corrections to the Callan-Gross ratio in heavy-flavor leptoproduction
are small, it makes sense to investigate in more detail the corresponding LO predictions.
In this Section, we derive compact low-$x$ approximation formulae for the ratio
$R_2(x,Q^2)=2xF_L/F_2$ at LO, which greatly simplify the extraction of the structure
function $F_2(x,Q^2)$ from measurements of the reduced cross section,
$\tilde{\sigma}(x,Q^{2})$, defined by Eqs.~(\ref{5}) and (\ref{6}). For this purpose, we
convolute the LO partonic cross sections given by Eqs.~(\ref{9}) and (\ref{10}) with the
low-$x$ asymptotics of the gluon PDF:
\begin{equation} \label{29}
g(x,Q^2)\stackrel{x\to 0}{\longrightarrow}\frac{1}{x^{1+\delta}}.
\end{equation}

The value of $\delta$ in Eq.~(\ref{29}) is a matter of discussion. The simplest choice,
$\delta =0$, leads to a non-singular behavior of the structure functions for $x\to 0$.
Another extreme value, $\delta =1/2$, historically originates from the BFKL resummation 
of the leading powers of $\ln(1/x)$ \cite{BFKL1}. In reality, $\delta$ is a function 
of $Q^2$ (for an experimental review, see Ref.~\cite{A-Vogt}). Theoretically, the $Q^2$ 
dependence of $\delta$ is calculated using the DGLAP evolution equations \cite{DGLAP}.

First, we calculate the LO hadron-level cross sections for both extreme cases,
$\delta =0$ and $1/2$. Our predictions for the quantity $R_2(x,Q^2)$ in the limit
of $x\to 0$ have the following form:
\begin{eqnarray}
R^{(0)}_2(Q^2)&=&\frac{2}{1+4\lambda}\,
\frac{1+6\lambda-4\lambda(1+3\lambda)J(\lambda)}{1+2(1-\lambda)J(\lambda)}, \label{30}\\
R^{(1/2)}_2(Q^2)&=&\frac{8}{1 + 4\lambda}\,\frac{ \left[ 3 + 4\lambda \left( 13
+ 32\lambda \right) \right] E(1/(1 + 4\lambda)) - 4\lambda \left( 9 +
32\lambda  \right) K(1/(1 + 4\lambda)) }{ \left( -37 + 72\lambda
\right)E(1/(1 + 4\lambda)) + 2\left( 23 - 36\lambda
\right)K(1/(1 + 4\lambda)) },\label{31}
\end{eqnarray}
where $\lambda$ is defined in Eq.~(\ref{12}),
\begin{equation} \label{32}
J(\lambda)=\frac{1}{\sqrt{1 + 4\lambda}}\ln \frac{\sqrt{1 + 4\lambda}+1}{\sqrt{1 +
4\lambda}-1},
\end{equation}
and the functions $K(y)$ and $E(y)$ are the complete elliptic integrals of the first and
second kinds defined as
\begin{equation} \label{33}
K(y)=\int\limits_0^1\frac{{\text d}t}{\sqrt{(1-t^2)(1-yt^2)}},
\qquad
E(y)=\int\limits_0^1{\text d}t \sqrt{\frac{1-yt^2}{1-t^2}}.
\end{equation}
The result in Eq.~(\ref{30}) was previously found in Ref.~\cite{kotikov}, where an
approximation to its NLO counterpart was also presented.

\begin{figure}
\begin{center}
\begin{tabular}{cc}
\mbox{\epsfig{file=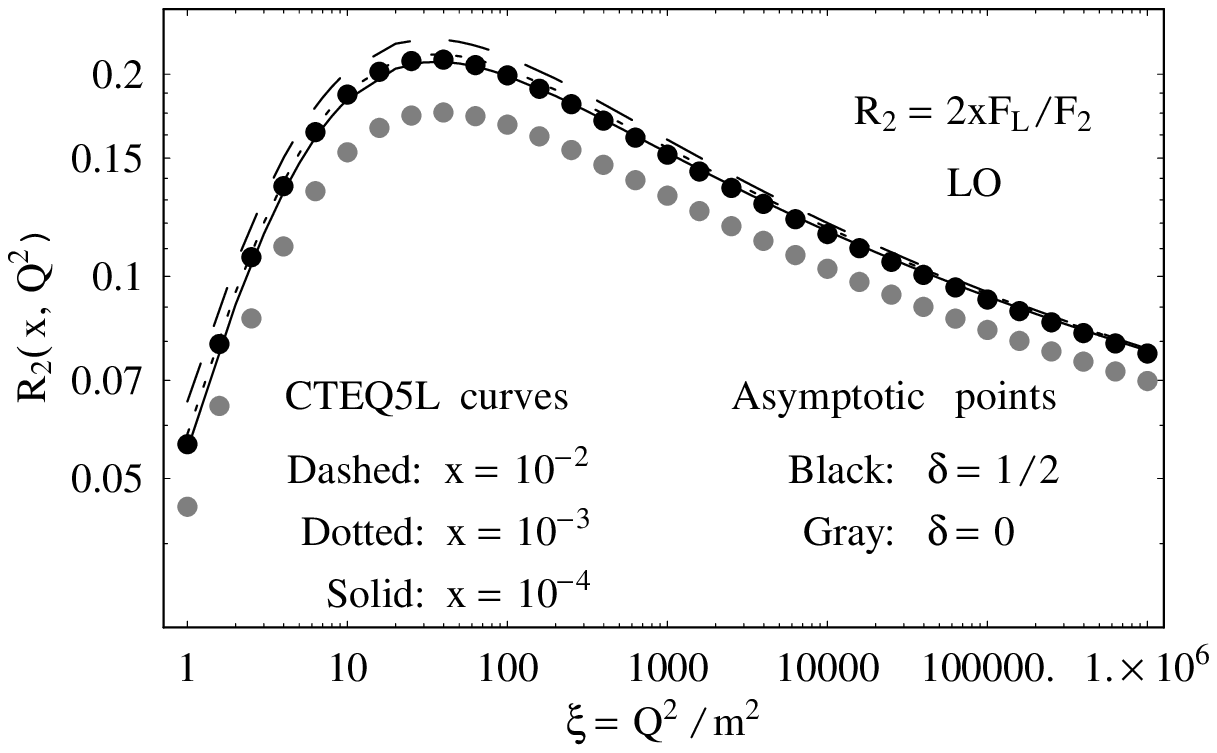,width=250pt}}
& \mbox{\epsfig{file=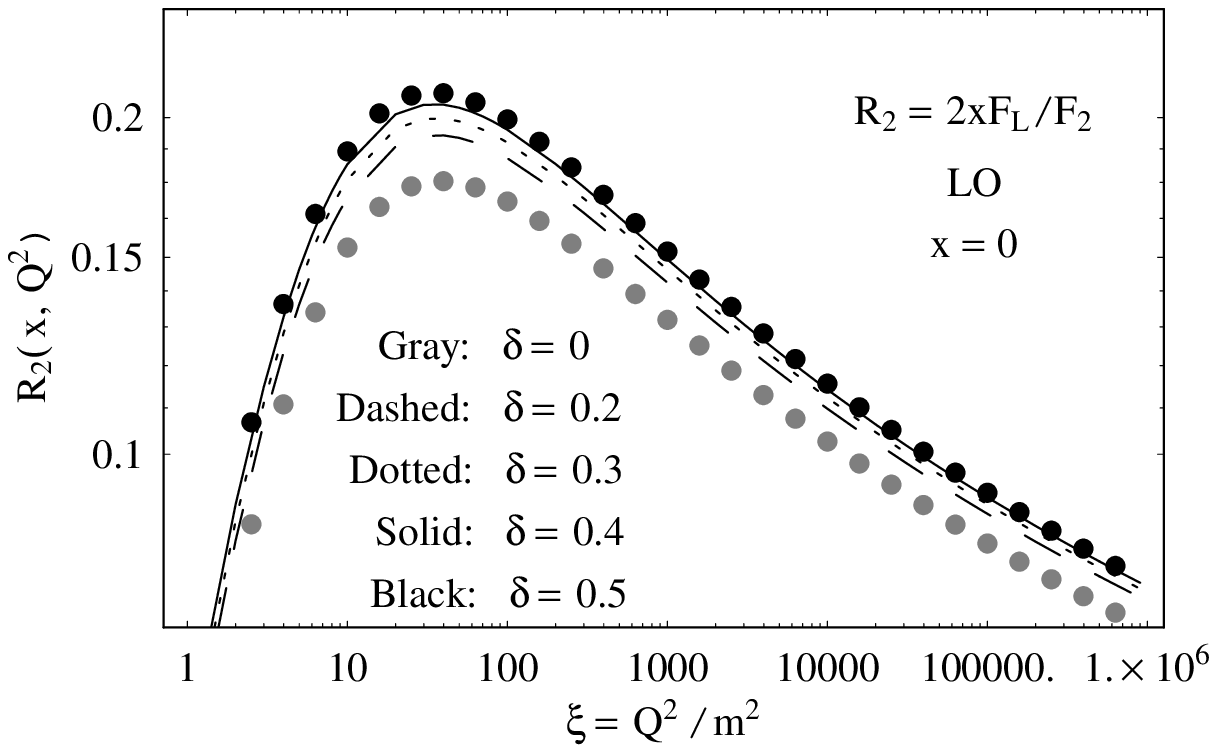,width=250pt}}\\
\end{tabular}
 \caption{\label{Fg.6}\small LO low-$x$ predictions for the ratio $R_2(x,Q^2)=2xF_L/F_2$
 in charm leptoproduction. \emph{Left panel:} Asymptotic ratios $R^{(0)}_2(Q^2)$
 (gray points) and $R^{(1/2)}_2(Q^2)$ (black points), as well as CTEQ5L
 predictions for $R_2(x,Q^2)$ at $x=10^{-2}$, $10^{-3}$ and $10^{-4}$.
 \emph{Right panel:}
 Asymptotic ratio $R^{(\delta )}_2(Q^2)$ at
 $\delta=0$, 0.2, 0.3, 0.4 and 0.5.}
\end{center}
\end{figure}
The left panel of Fig.~\ref{Fg.6} shows the ratios $R^{(0)}_2(Q^2)$ and
$R^{(1/2)}_2(Q^2)$ as functions of $\xi$. One can see that the difference between these
quantities varies slowly from $20\%$ at low $Q^2$ to $10\%$ at high $Q^2$. For
comparison, also the LO results for $R_2(x,Q^2)$ calculated at several values of $x$
using the CTEQ5L gluon PDF \cite{CTEQ5} are shown. We observe that, for $x\to 0$, the
CTEQ5L predictions converge to the function $R^{(1/2)}_2(Q^2)$ practically in the entire
region of $Q^2$. We have verified that the same situation takes also place for all other
LO and NLO CTEQ PDF versions \cite{CTEQ6,CTEQ5}.

Next, we derive an analytic low-$x$ formula for the ratio $R^{(\delta)}_2(x,Q^2)$
with arbitrary values of $\delta$, in terms of the Gauss hypergeometric function.
Our result has the following form:
\begin{equation} \label{34}
R_{2}^{(\delta )}(Q^2)=4\frac{\frac{2+\delta }{3+\delta }\Phi
\left( 1+\delta ,\frac{1}{1+4\lambda }\right) -\left( 1+4\lambda \right)
\Phi \left( 2+\delta ,\frac{1}{1+4\lambda }\right) }{\left[ 1+\frac{%
\delta \left( 1-\delta ^{2}\right) }{\left( 2+\delta \right) \left( 3+\delta \right)
}\right] \Phi \left( \delta ,\frac{1}{1+4\lambda }\right) -\left( 1+4\lambda \right)
\left( 4-\delta -\frac{10}{3+\delta }\right) \Phi \left( 1+\delta ,\frac{1}{1+4\lambda
}\right) },
\end{equation}
where the function $\Phi (r,z)$ is defined as
\begin{equation} \label{35}
\Phi \left( r,z\right) =\frac{z^{1+r}}{1+r}\,\frac{\Gamma \left( 1/2\right) \Gamma \left(
1+r\right) }{\Gamma \left( 3/2+r\right) }\,{}_{2}F_{1}\left(
\frac{1}{2},1+r;\frac{3}{2}+r;z\right) .
\end{equation}
The hypergeometric function ${}_{2}F_{1}(a,b;c;z)$ has the following series expansion:
\begin{equation}
{}_{2}F_{1}\left( a,b;c;z\right)=\frac{\Gamma \left( c\right) }{\Gamma \left( a\right)
\Gamma \left( b\right)}\sum\limits_{n=0}^{\infty }\frac{\Gamma
\left( a+n\right) \Gamma \left( b+n\right) }{\Gamma \left( c+n\right) }\frac{%
z^{n}}{n!}. \label{36}
\end{equation}

In the right panel of Fig.~\ref{Fg.6}, the $\delta$ dependence of the asymptotic ratio
$R^{(\delta)}_2(Q^2)$ is investigated. One can see that the ratio $R^{(\delta)}_2(Q^2)$
rapidly converges to the function $R^{(1/2)}_2(Q^2)$ for $\delta > 0.2$. In particular,
the relative difference between $R^{(0.5)}_2(Q^2)$ and $R^{(0.3)}_2(Q^2)$ varies slowly
from $6\%$ at low $Q^2$ to $2\%$ at high $Q^2$. 

As mentioned above, the $Q^2$ dependence of the parameter $\delta$ is determined with the 
help of the DGLAP evolution. However, our analysis shows that hadron-level predictions for 
$R_2(x\to 0,Q^2)$ depend weakly on $\delta$ practically in the entire region of $Q^2$ 
for $0.2< \delta < 0.9$. For this reason, our simple formula (\ref{31}) with $\delta =1/2$ 
(i.e., without any evolution) describes with good accuracy the low-$x$ CTEQ results for 
$R_2(x,Q^2)$. We conclude that the hadron-level 
predictions for $R_2(x\to 0,Q^2)$ are stable not only under the NLO corrections to 
the partonic cross sections, but also under the DGLAP evolution of the gluon PDF.

Finally, we use the analytic expressions (\ref{30}), (\ref{31}) and (\ref{34}) for the
extraction of the structure functions $F_2^c(x,Q^2)$ and $F_2^b(x,Q^2)$ from the HERA
measurements of the reduced cross sections $\tilde{\sigma}^c(x,Q^2)$ and
$\tilde{\sigma}^b(x,Q^2)$, respectively. The results of our analysis of the HERA data on
the charm and bottom electroproduction are collected in Tables~\ref{tab1} and \ref{tab2},
respectively. In our calculations, the values $m_c=1.3$~GeV and $m_b=4.3$~GeV for the
charm and bottom quark masses are used. The LO predictions, $F_2(\mathrm{LO})$, for the
cases of $\delta=0.5$, $0.3$ and $0$ are presented and compared with the NLO values,
$F_2(\mathrm{NLO})$, obtained in the H1 analysis \cite{H1HERA1,H1HERA2}. One can see that
all the considered LO predictions agree with the NLO results with an accuracy better than
1\%. This is because the contributions of the longitudinal structure functions,
$F_L^c(x,Q^2)$ and $F_L^b(x,Q^2)$, to the reduced cross sections,
$\tilde{\sigma}^c(x,Q^2)$ and $\tilde{\sigma}^b(x,Q^2)$, are small, less than 5\%, in the
kinematic range of the HERA H1 experiment.

\begin{table}[h]
\caption{\label{tab1} Values of $F_2^c(x,Q^2)$ extracted from the HERA measurements of
$\tilde{\sigma}^c(x,Q^{2})$ at low \cite{H1HERA2} and high \cite{H1HERA1} $Q^2$ (in
GeV$^2$) for various values of $x$ (in units of $10^{-3}$). The
NLO H1 results \cite{H1HERA1,H1HERA2} are compared with the LO predictions corresponding to
the cases of $\delta =0.5$, $0.3$ and $0$.}
\begin{tabular}{||ccc||cc||cccc||}
\hline
 $\quad Q^2 \quad$ & $x$ & ~$\quad y\quad$~ & ~$\quad \tilde{\sigma}^c \quad$~ &
  ~Error~ &
$\qquad F_2^c$(NLO)$\qquad$ & $ F_2^c$(LO) &$\qquad$ $ F_2^c$(LO) $\qquad$&  $F_2^c$(LO) \\
 (GeV$^2$) & $(\times 10^{-3})$  &  &  & (\%) & H1 & $\delta=0.5$ & $\delta=0.3$ & $\delta=0$\\
\hline
  \hline
 12 & 0.197 & 0.600 & 0.412 & 18 & $0.435\pm0.078$ & $0.435\pm0.078$ & $0.434\pm0.078$ & $0.431\pm0.077$ \\
 12 & 0.800 & 0.148 & 0.185 & 13 & $0.186\pm0.024$ & $0.185\pm0.024$ & $0.185\pm0.024$ & $0.185\pm0.024$ \\
 25 & 0.500 & 0.492 & 0.318 & 13 & $0.331\pm0.043$ & $0.331\pm0.043$ & $0.330\pm0.043$ & $0.328\pm0.043$ \\
 25 & 2.000 & 0.123 & 0.212 & 10 & $0.212\pm0.021$ & $0.212\pm0.021$ & $0.212\pm0.021$ & $0.212\pm0.021$ \\
 60 & 2.000 & 0.295 & 0.364 & 10 & $0.369\pm0.040$ & $0.369\pm0.040$ & $0.368\pm0.040$ & $0.368\pm0.040$ \\
 60 & 5.000 & 0.118 & 0.200 & 12 & $0.201\pm0.024$ & $0.200\pm0.024$ & $0.200\pm0.024$ & $0.200\pm0.024$ \\
200 & 0.500 & 0.394 & 0.197 & 23 & $0.202\pm0.046$ & $0.202\pm0.046$ & $0.202\pm0.046$ & $0.201\pm0.046$ \\
200 & 1.300 & 0.151 & 0.130 & 24 & $0.131\pm0.032$ & $0.130\pm0.031$ & $0.130\pm0.031$ & $0.130\pm0.031$ \\
650 & 1.300 & 0.492 & 0.206 & 27 & $0.213\pm0.057$ & $0.213\pm0.057$ & $0.213\pm0.057$ & $0.212\pm0.057$ \\
650 & 3.200 & 0.200 & 0.091 & 31 & $0.092\pm0.028$ & $0.091\pm0.028$ & $0.091\pm0.028$ & $0.091\pm0.028$ \\
\hline
\end{tabular}
\end{table}
\begin{table}[h]
\caption{\label{tab2} Values of $F_2^b(x,Q^2)$ extracted from the HERA measurements of
$\tilde{\sigma}^b(x,Q^{2})$ at low \cite{H1HERA2} and high \cite{H1HERA1} $Q^2$ (in
GeV$^2$) for various values of $x$ (in units of $10^{-3}$). The
NLO H1 results \cite{H1HERA1,H1HERA2} are compared with the LO predictions corresponding to
the cases of $\delta =0.5$, $0.3$ and $0$.}
\begin{tabular}{||ccc||cc||cccc||}
\hline
 $\quad Q^2 \quad$ & $x$ & ~$\quad y\quad$~ & ~$\quad \tilde{\sigma}^b \quad$~ &
  ~Error~ &
$\qquad F_2^b$(NLO)$\qquad$ & $ F_2^b$(LO) &$\qquad$ $ F_2^c$(LO) $\qquad$&  $F_2^b$(LO)\\
 (GeV$^2$) & $(\times 10^{-3})$  &  &  & (\%) & H1 & $\delta=0.5$ & $\delta=0.3$ & $\delta=0$\\
\hline
  \hline
 12 & 0.197 & 0.600 & 0.0045 & 60 & $0.0045\pm0.0027$ & $0.0046\pm0.0027$ & $0.0046\pm0.0027$ & $0.0046\pm0.0027$ \\
 12 & 0.800 & 0.148 & 0.0048 & 45 & $0.0048\pm0.0022$ & $0.0048\pm0.0022$ & $0.0048\pm0.0022$ & $0.0048\pm0.0022$ \\
 25 & 0.500 & 0.492 & 0.0122 & 31 & $0.0123\pm0.0038$ & $0.0124\pm0.0038$ & $0.0124\pm0.0038$ & $0.0123\pm0.0038$ \\
 25 & 2.000 & 0.123 & 0.0061 & 39 & $0.0061\pm0.0024$ & $0.0061\pm0.0024$ & $0.0061\pm0.0024$ & $0.0061\pm0.0024$ \\
 60 & 2.000 & 0.295 & 0.0189 & 29 & $0.0190\pm0.0055$ & $0.0190\pm0.0055$ & $0.0190\pm0.0055$ & $0.0190\pm0.0055$ \\
 60 & 5.000 & 0.118 & 0.0130 & 36 & $0.0130\pm0.0047$ & $0.0130\pm0.0047$ & $0.0130\pm0.0047$ & $0.0130\pm0.0047$ \\
200 & 0.500 & 0.394 & 0.0393 & 31 & $0.0413\pm0.0128$ & $0.0402\pm0.0125$ & $0.0401\pm0.0125$ & $0.0400\pm0.0124$ \\
200 & 1.300 & 0.151 & 0.0212 & 38 & $0.0214\pm0.0081$ & $0.0213\pm0.0081$ & $0.0213\pm0.0081$ & $0.0212\pm0.0081$ \\
650 & 1.300 & 0.492 & 0.0230 & 51 & $0.0243\pm0.0124$ & $0.0240\pm0.0122$ & $0.0239\pm0.0122$ & $0.0238\pm0.0121$ \\
650 & 3.200 & 0.200 & 0.0124 & 44 & $0.0125\pm0.0055$ & $0.0125\pm0.0055$ & $0.0125\pm0.0055$ & $0.0125\pm0.0055$ \\
\hline
\end{tabular}
\end{table}

\section{Conclusion}

We conclude by summarizing our main observations. In the present paper, we studied
the radiative corrections to the Callan-Gross ratio $R(x,Q^{2})$ in heavy-quark
leptoproduction. We considered the exact NLO results at low and moderate
$Q^2\lesssim m^2$, asymptotic NLO predictions at high $Q^2\gg m^2$, and both NLO and NNLO
soft-gluon (or threshold) corrections at large Bjorken $x$. It turned out that large
(especially, at non-small $x$) radiative corrections to the structure functions
$F_T(x,Q^2)$ and
$F_L(x,Q^2)$ cancel each other in their ratio $R(x,Q^2)=F_L/F_T$ with good accuracy.
As a result, the
NLO contributions to the ratio $R(x,Q^{2})$ are less than $10\%$ in a wide region of the
variables $x$ and $Q^2$. Our analysis also shows that the NLO predictions for $R(x,Q^2)$
are
sufficiently insensitive (to within ten percent) to standard uncertainties in the QCD
input parameters. We conclude that, unlike the production cross sections, the
Callan-Gross ratio in heavy-quark leptoproduction is quantitatively well
defined in pQCD. Measurements of the quantity $R(x,Q^2)$ in charm and bottom
leptoproduction
would provide a good test of the conventional parton model based on pQCD.

Concerning the experimental aspects, we propose to exploit the observed perturbative
stability of the Callan-Gross ratio in the extraction of the structure functions
$F_2^c(x,Q^2)$ and $F_2^b(x,Q^2)$ from the corresponding reduced cross sections.
For this purpose, we provided
compact LO hadron-level formulae for the ratio $R_{2}(x,Q^2)=2xF_L/F_2=R/(1+R)$ in the
limit $x\to 0$. We demonstrated that these analytic expressions simplify the extraction 
of $F_{2}(x,Q^2)$ without affecting the accuracy of the result in practice. In particular, 
our LO formula for $R_{2}(x,Q^2)$ with $\delta=1/2$ usefully reproduces the results for 
$F_2^c(x,Q^2)$ and $F_2^b(x,Q^2)$ obtained by the H1 Collaboration \cite{H1HERA1,H1HERA2} 
with the help of the more cumbersome NLO evaluation of $F_{L}(x,Q^2)$.

In this paper, we investigated the contribution to $R(x,Q^2)$ from the dominant mechanism,
photon-gluon fusion, within the fixed-flavor-number scheme.
To take into account the photon-heavy-quark scattering component, one should
adopt the variable-flavor-number scheme, which allows one to resum potentially large
mass logarithms of the type $\alpha_{s}\ln\left( Q^{2}/m^{2}\right)$, whose contribution
dominates at $Q^{2}\gg m^2$. Some recent developments concerning this scheme
may be found in Ref.~\cite{SACOT}.
The variable-flavor-number-scheme predictions for the Callan-Gross ratio as well as the
possibility to discriminate experimentally between photon-gluon fusion and
quark-scattering contributions to $R(x,Q^2)$ will be considered in a forthcoming publication.

\begin{acknowledgments}
N.Ya.I. thanks S.J. Brodsky for drawing his attention to the problem considered in this
paper.
We are grateful A.V. Kotikov for interesting and useful discussions.
This work was supported in part by BMBF Grant No.\ 05~HT6GUA.

\end{acknowledgments}

\end{document}